\documentclass[12pt,  leqno, ]{article}
\usepackage[dvips]{graphicx}
\usepackage{float}
\usepackage{latexsym}

\usepackage{authblk}

\usepackage{amsmath,mathrsfs,bm,amssymb,color}
\usepackage{graphicx}
\usepackage{wrapfig}

\newtheorem{theorem}{Theorem}[section]
\newtheorem{proposition}[theorem]{Proposition}
\newtheorem{lemma}[theorem]{Lemma}
\newtheorem{corollary}[theorem]{Corollary}

\newtheorem{example}[theorem]{Example}
\newtheorem{remark}[theorem]{Remark}

\newtheorem{assumption}[theorem]{Assumption}

\def\bbbone{{\mathchoice {\rm 1\mskip-4mu l} {\rm 1\mskip-4mu l}
{\rm 1\mskip-4.5mu l} {\rm 1\mskip-5mu l}}}
\def\one{\bbbone}

\newcommand{\proof}{{\noindent \it Proof:\ }}
\newcommand{\qed}{\hfill $\Box$\par\medskip}

\newcommand{\bi}{
\begin{itemize}
}
\newcommand{\ei}{\end{itemize}}

\newcommand{\KK}{{\mathcal K}}

\renewcommand{\d}{\displaystyle}

\newcommand{\BR}
{{{\mathbb R}^d}}
\newcommand{\RR}{{{\mathbb R}}}

\newcommand{\jjj}{\sum_{j=1}^{d-1}}

\newcommand{\LR}{{L^2({\mathbb R}^d)}}

\newcommand{\s}{\sigma}

\newcommand{\fff}{{\ms F}}

\newcommand{\f}{^{-1}}

\newcommand{\ov}[1]{\overline{#1}}

\newcommand{\add}{a^{\ast}}

\newcommand{\mmm}{\lk
\frac{d-1}{d}\rk}

\newcommand{\lk}{\left(}
\newcommand{\rk}{\right)}

\newcommand{\ass}{a^{\sharp}}

\newcommand{\han}{{1/2}}

\newcommand{\hhh}{{\ms H}}
\newcommand{\ms}{\mathscr}

\newcommand{\hf}{{H_{\rm f}}}

\newcommand{\mass}{m_{\rm eff}(\alpha)}
\newcommand{\is}{\inf\!\s}

\newcommand{\vp}{\hat{\varphi}}

\newcommand{\eq}[1]{\begin{equation}
\label{#1}}
\newcommand{\en}{\end{equation}}

\newcommand{\eqn}
{\begin{eqnarray*}
}
\newcommand{\enn}{\end{eqnarray*}}

\newcommand{\bt}[1]{\begin{theorem}
\label{#1}}
\newcommand{\et}{\end{theorem}}
\newcommand{\bl}[1]{\begin{lemma}
\label{#1}}
\newcommand{\enl}{\end{lemma}}

\newcommand{\bc}[1]{\begin{corollary}
\label{#1}}
\newcommand{\ec}{\end{corollary}}

\newcommand{\kak}[1]{(\ref{#1})}

\newcommand{\hp}{H_{\rm p}}

\newcommand{\hz}{-\frac{1}{2\mass}\Delta}

\makeatletter
\@addtoreset{equation}{section}
\makeatother

\begin{document}
\title
{The No-Binding Regime
of the Pauli-Fierz Model\vspace{20pt}}

\author[,1]{Fumio Hiroshima}
\author[,2]{Herbert Spohn}
\author[,3]{Akito Suzuki\vspace{16pt}}
 \affil[1]{Faculty of Mathematics, Kyushu University, Fukuoka, 819-0395, Japan}
\affil[2]{
Zentrum Mathematik and Physik Department, TU M\"unchen, D-80290,
M\"unchen, Germany }
\affil[3]{Department of Mathematics, Faculty of Engineering, \\ Shinshu University, Nagano, 380-8553, Japan}
\date{\today}
\maketitle


\begin{abstract}
The Pauli-Fierz model $H(\alpha)$ in nonrelativistic quantum electrodynamics is considered.
The external potential $V$
is sufficiently shallow
and the dipole approximation is assumed.
It is proven that
there exist constants
$0<\alpha_-< \alpha_+$ such that
$H(\alpha)$ has no ground state for
$|\alpha|<\alpha_-$,
which complements an earlier result stating that
there is a ground state for $|\alpha| > \alpha_+$. We develop a
suitable extension of the Birman-Schwinger argument.
Moreover for any given $\delta>0$
examples of potentials $V$ are provided such that $\alpha_+-\alpha_-<\delta$.
\end{abstract}


\setlength{\baselineskip}{14pt}
\newpage
\section{Introduction}
Let us consider a quantum particle in an external potential described by the Schr\"odinger operator
\begin{equation}
\label{schr11}
\hp(m) = -\frac{1}{2m} \Delta + V(x)
\end{equation}
acting on $L^2(\RR^d)$. If the potential $V$ is short ranged and attractive and if the dimension $d \geq 3$, then there is a transition from unbinding to binding as the mass $m$ is increased.
More precisely,
there is some critical mass, $m_{\mathrm{c}}$,
such that
$\hp(m)$ has no ground state for
$0 < m < m_{\mathrm{c}}$ and a unique ground state for $m_{\mathrm{c}} < m$.
In fact, the critical mass is given by
$$
\frac{1}{2m_{\mathrm{c}}}=
\left\|
|V|^\han \lk
-\Delta\rk\f |V|^\han
\right\|,$$
see  Lemma \ref{schr}.
We now couple $\hp(m)$ to the quantized electromagnetic field
with coupling strength $\alpha \geq 0$. The corresponding Hamiltonian is denoted by $H(\alpha)$.
On a heuristic level, through the dressing by photons
the particle becomes effectively more heavy,
which means that the critical mass $c_0 \alpha^2(\alpha)$
should be decreasing as a function of $\alpha$
with $m_{\mathrm{c}}(0) = m_{\mathrm{c}}$.
In particular, if $m < m_{\mathrm{c}}$, then there should be an unbinding-binding transition as the coupling
$\alpha$ is increased. This phenomenon has been baptized  {\it enhanced binding}  and
has been studied for a variety of models
by several authors \cite{arka,bv04,hvv,hhs,hs01,hs08}. In case $m>m_{\mathrm{c}}$
more general techniques are available  and the existence of a unique ground state
for the full Hamiltonian is proven in   \cite{AH,BFS,GLL,ll03,G,Sp}.
\begin{figure}[h]
 \begin{center}
   \includegraphics[width=70mm]{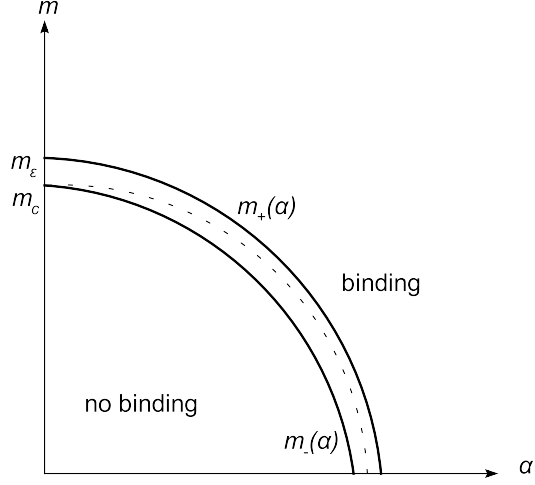}
 \caption{Upper and lower bounds on the critical mass $m_\mathrm{c}(\alpha)$. The dashed line indicates $m_\mathrm{c}(\alpha)$}
 \end{center}
\end{figure}

The heuristic picture also asserts that the full hamiltonian has a regime of couplings with no ground state.
This property is more difficult to establish and the only result we are aware of is proved by Benguria and Vougalter \cite{bv04}. In essence they establish that the line $m_\mathrm{c}(\alpha)$ is continuous as
$ \alpha \to 0$. (In fact, they use the strength of the potential as parameter). From this it follows that the no binding regime cannot be empty. In our paper, as in \cite{hs01}, we will use the dipole approximation for simplicity,
but provide a fairly explicit bound on the critical mass. In the dipole approximation the effective mass
$m_\mathrm{eff}(\alpha) = m +c_0 \alpha^2$ with some explicitly computable coefficient $c_0$, see Eq. (\ref{2.10})
below. Thus the most basic guess for $m_{\mathrm{c}}(\alpha)$ would be $m_{\mathrm{c}}(\alpha) + c_0 \alpha^2
= m_{\mathrm{c}}$. The corresponding curve is displayed in Fig. 1. In fact the guess turns out to be a lower bound on the true $m_{\mathrm{c}}(\alpha)$. We will complement our lower bound with an upper bound of the same qualitative form.

The unbinding for the Schr\"odinger operator $\hp(m)$
is proven by the Birman-Schwinger principle.
Formally one has
 $$ \hp(m) =
 \frac{1}{2m}
 (-\Delta)^\han
 \big(\one +
 2m (-\Delta)^{-\han}
 V
 (-\Delta)^{-\han}
  \big)
 (-\Delta)^\han.
 $$
 If $m$ is sufficiently small, then
 $
 2m(-\Delta)^{-\han}
 V
 (-\Delta)^{-\han}
 $
 is a strict contraction.
 Hence
the  operator $\one +
 2m (-\Delta)^{-\han}
 V
 (-\Delta)^{-\han}
 $ has a bounded inverse
 and $\hp(m)$ has no eigenvalue in $(-\infty,0]$.
More precisely the Birman-Schwinger principle
states that
\eq{a6}
{\rm dim}\one_{[\frac{1}{2m},\infty)
}
(V^\han (-\Delta)\f V^\han)
\geq
{\rm dim}
\one_{(-\infty,0]}
(\hp(m)).
\en
For small $m$ the left hand side equals $0$ and thus
$\hp(m)$ has no eigenvalues in $(-\infty,0]$.

Our approach will be to generalize
\kak{a6} to the Pauli-Fierz model of non-relativistic quantum electrodynamics.
The Pauli-Fierz Hamiltonian $H(\alpha)$
is defined on the Hilbert space $\hhh=\LR\otimes \ms F$, where $\ms F$ denotes
the boson Fock space.
Transforming $H(\alpha)$ unitarily by $U$
one arrives at
\begin{eqnarray}\label{a7}
&&U\f H(\alpha)U =
H_0(\alpha) + W +g
\end{eqnarray}
as the sum of the free Hamiltonian
\eq{aa7}
 H_0(\alpha) = -\frac{1}{2\mass}\Delta\otimes \one+\one\otimes\hf,
\en
involving the effective mass of the dressed particle and the Hamiltonian  $\hf$ of the free boson field,
the transformed interaction
\eq{aa8}
W = T\f (V\otimes \one) T,
\en
and the global energy shift $g$.
 $\mass$ is an increasing function of $\alpha$.
We will show that  \kak{a7}
has no ground state
for sufficiently small $|\alpha|$ by means of
a Birman-Schwinger type argument such as \kak{a6}.
In combination with the results obtained in \cite{hs01}
we provide examples of external potentials $V$ such that for some given $\delta>0$
there exist two constants $0<\alpha_-<\alpha_+$ satisfying
\eq{s7}
\delta>\alpha_+-\alpha_->0
\en
and
$H(\alpha)$ has no ground state for $|\alpha|<\alpha_-$ but
has a ground state for $|\alpha|>\alpha_+$.

Our paper is organized as follows.
In Section 2 we define the Pauli-Fierz model and
in Section 3 we prove the absence of ground states.
Section 4  lists examples of
external potentials exhibiting the unbinding-binding  transition.


\section{The Pauli-Fierz Hamiltonian}
 We assume a space dimension $d\geq 3$ throughout, and
 take the natural unit:
the velocity of light $c=1$ and the Planck constant divided $2\pi$, $\hbar=1$.
The Hilbert space $\hhh$ for the Pauli-Fierz Hamiltonian is
given by $$\hhh=\LR\otimes\fff, $$
where
$$\fff=\bigoplus_{n=0}^\infty
   \left[
\otimes_s^n (\oplus^{d-1}\LR)\right]$$
denotes the boson Fock space over the $(d-1)$-fold direct sum  $\oplus^{d-1}\LR$.
Let $\Omega=\{1,0,0,...\}\in\fff$ denote the Fock vacuum.
The creation operator
and the annihilation operator
are
denoted by
$\add(f,j)$ and $a(f,j)$, $j=1,\ldots,d-1$, $f\in\LR$, respectively,
and they
 satisfy the canonical commutation relations
 \begin{eqnarray*}
 [a(f,j),\add(g,j')]=\delta_{jj'}(f,g)\one,  \quad  [a(f,j),a(g,j')]=0=
 [\add(f,j),\add(g,j')]
 \end{eqnarray*}
with $(f,g)$ the scalar product on $\LR$.
We write
\eq{s6}
\ass(f,j)=\int \ass(k,j) f(k) dk,\quad \ass=a, \add,
\en
 The energy of a single photon with momentum $k\in\BR$
is
\eq{10}
\omega(k)=|k|.
\en
 The free Hamiltonian on $\fff$
is then given by
\eq{s8}
\hf= \jjj \int\omega(k)\add(k,j)a(k,j) dk .
\en
Note that
$\s(\hf)=[0,\infty)$,
and $\s_{\rm p}(\hf)=\{0\}$.
$\{0\}$ is a simple eigenvalue of $\hf$
and
$\hf\Omega=0$.

Next we introduce the quantized radiation field.
The
$d$-dimensional polarization vectors are denoted by
 $e_j(k)\in\BR$,
$j=1,\ldots,d-1$,
which satisfy
$e_i(k)\cdot e_j(k)=\delta_{ij}$ and
 $e_j(k)\cdot k=0$ almost everywhere on $\BR$.
 The quantized vector potential then reads
\eq{b1}
A(x)
=\jjj \int\frac{1}{\sqrt{2\omega(k)}}
e_j(k)\big(
\vp(k) a^\ast(k,j){\mathrm e}^{-ikx}
+\vp(-k) a(k,j){\mathrm e}^{ikx}
\big) dk
\en
for $x\in\BR$ with ultraviolet cutoff $\vp$.
Conditions imposed on $\vp$ will be supplied later.
Assuming that $V$ is centered, in the dipole approximation
 $A(x)$ is replaced by $A(0)$.  We set $A=A(0)$.
 The Pauli-Fierz Hamiltonian $H(\alpha)$ in the dipole approximation
 is then given  by
\eq{s18}
H(\alpha)=\frac{1}{2m }\lk p\otimes \one-\alpha \one \otimes A\rk^2+V\otimes \one+\one \otimes \hf,
\en
where $\alpha\in\RR$ is the coupling constant,
$V$ the external potential, and $p=(-i\partial_1,...,-i\partial_d)$ the momentum operator.
For notational convenience we omit the tensor notation $\otimes$ in what follows.
\begin{assumption}
\label{kimi}
Suppose that $V$ is relatively bounded
with respect to
$\d-\frac{1}{2m}\Delta$
with a relative bound strictly smaller than one,
and
\eq{b2}
\vp/\omega\in\LR,\quad \sqrt\omega\vp\in\LR.
\en
\end{assumption}
By this assumption
$H(\alpha)$ is self-adjoint on $D(-\Delta)\cap D(\hf)$ and bounded
below for arbitrary $\alpha\in\RR$ \cite{ar1,ar4}.
We need in addition some technical assumptions on $\vp$
which are introduced in
\cite[Definition 2.2]{hs01}.
We list them as
\begin{assumption}
\label{1v}
The ultraviolet cutoff $\vp$ satisfies (1)-(4) below.
\bi
\item[(1)]$\vp/\omega^{3/2}\in\LR$;
\item[(2)]
$\vp$ is rotation invariant, i.e.  $\vp(k)=\chi(|k|)$ with some real-valued function
$\chi$ on $[0,\infty)$;
and
  $\rho(s)=|\chi(\sqrt{s})|^2s^{(d-2)/2}\in L^\epsilon([0,\infty),ds)$ for some
$1<\epsilon$,and there exists $0<\beta<1$ such that
$|\rho(s+h)-\rho(s)|\leq K|h|^\beta$ for all $s$ and $0<h\leq 1$ with some constant $K$;
\item[(3)]
  $\|\vp\omega^{(d-1) /2}\|_\infty<\infty$;
\item[(4)]
$\vp(k)\not=0$ for $k\not=0$.
\ei
\end{assumption}
The Hamiltonian $H(\alpha)$ with $V=0$ is quadratic and can therefore be diagonalized explicitly,
which is carried out in \cite{ar4,hs01}.
Assumption \ref{1v} ensures
the existence of a unitary operator diagonalizing $H(\alpha)$.

Let $$D_+(s)=m-\alpha^2\frac{d-1}{d}\int\frac{|\vp(k)|^2}{s-\omega(k)^2+i0}dk,\quad s\geq0.$$
We see that $D_+(0)=m+\alpha^2\frac{d-1}{d}\|\vp/\omega\|^2>0$ and the imaginary part of $D_+(s)$ is
$\alpha^2\frac{d-1}{d}\pi S_{d-1}\rho(s)\not=0$ for $s\not=0$,
where $\rho$ is defined in (2) of Assumption \ref{1v} and $S_{d-1}$ is
the volume of the $d-1$ dimensional unit sphere, and the real part of $D_+(s)$ satisfies that $\lim_{s\to\infty}\Re D_+(s)=m>0$.
These properties follows from Assumption \ref{1v}.
In particular
\eq{kansai}
\inf_{s\geq 0}|D_+(s)|>0.
\en
Define \eq{aa11}
\Lambda_j^\mu(k)=\frac{e_j^\mu(k)\vp(k)}{\omega^{3/2}(k){D_+(\omega^2(k))}}.
\en
Then
$\|\Lambda_j^\mu \|\leq C
\|\vp/\omega^{3/2}\|$
for some constant $C$.

\begin{proposition}
\label{kansai1}
Under the assumptions \ref{kimi} and  \ref{1v},
for each $\alpha\in\RR$,
there exist
unitary operators $U$ and $T$
on  $\hhh$ such that
both  map $D(-\Delta)\cap
D(\hf)$ onto itself and
\eq{ni}
U\f H(\alpha)U=-\frac{1}{2\mass}\Delta
  +\hf+T\f V T +g,
\en
where $\mass $ and $g$ are constants given by
\begin{eqnarray}\label{2.10}
&&
\mass=m+\alpha^2 \mmm \|\vp/{\omega}\|^2,\\
&&
g=\frac{d}{2\pi}
\int_{-\infty}^\infty
\frac{t^2 \alpha^2   \mmm\|\vp/(t^2+\omega^2)\|^2}
{m+\alpha^2   \mmm\|\vp/\sqrt{t^2+\omega^2}\|^2}dt.
\end{eqnarray}
Here $U$ is defined in (4.29) of \cite{hs01} and
$T$  by \eq{a3}
T=\exp\left(-i\frac{\alpha}{\mass}p\cdot \phi\right),
\en
where $\phi=(\phi_1,...,\phi_d)$ is the vector field
$$\phi_\mu=\frac{1}{\sqrt 2}  \jjj \int
\lk \ov{\Lambda_j^\mu(k)}
 \add  (k,j)+ {\Lambda_j^\mu (k)} a(k,j)\rk dk.$$
\end{proposition}
\proof See \cite[Appendix]{hs01}.
\qed


\section{The Birman-Schwinger principle}
\subsection
{The case of Schr\"odinger operators}
Let $h_0 = - \frac{1}{2} \Delta$.
We assume that $V \in L^1_{\rm loc}(\RR^d)$ and $V$ is relatively form-bounded
with respect to $h_0$ with relative bound $a<1$, i.e.,
$D(|V|^{1/2}) \supset D(h_0^{1/2})$ and
\begin{equation}
\label{1336}
| |V|^{1/2} \varphi \|^2 \leq a \| h_0^{1/2}\varphi \|^2 + b \| \varphi\|^2,
	\quad \varphi \in D(h_0^{1/2}),
\end{equation}
with some $b>0$.
Then the operators
\eq{yu1}
 R_E = \left(h_0 -E\right)^{-1/2} |V|^{1/2},
	\quad \quad E < 0,
\en
are densely defined.
From \eqref{1336} it follows that
$R_E^* = |V|^{1/2}(h_0-E)^{-1/2}$ is bounded and thus
$R_E$ is closable.
We denote its closure by
the same symbol.
Let
\eq{s9}
 K_E = R_E^* R_E.
 \en
 Then $K_E$ ($E<0)$ is a bounded, positive self-adjoint operator
and it holds
\[ K_E f = |V|^{1/2}\left( h_0 -E \right)^{-1}|V|^{1/2} f, \quad f \in C_0^\infty(\RR^d). \]
Now let us consider the case $E=0$.
Let
\eq{yu2}
R_0 = h_0^{-1/2} |V|^{1/2}.
\en
The self-adjoint operator $h_0^{-1/2}$ has the integral kernel
\[ h_0^{-1/2}(x,y) =  \frac{a_d}{|x-y|^{d-1}}, \quad d\geq3,\]
where $a_d = \sqrt{2}\pi^{(d-1)/2}/\Gamma((d-1)/2)$
and $\Gamma(\cdot)$ the Gamma function.
It holds that
\[ \left|(h_0^{-1/2}g, |V|^{1/2} f) \right| \leq a_d \|g\|_2 \||V|^{1/2}f\|_{2d/(d+2)} \]
for $f, g \in C_0^\infty(\RR^3)$
by the Hardy-Littlewood-Sobolev inequality.
Since $f\in C_0^\infty(\RR^3)$ and $V \in L^1_{\rm loc}(\RR^3)$, one concludes
$\||V|^{1/2}f\|_{2d/(d+2)}
< \infty$.
Thus
$|V|^{1/2} f \in D(h_0^{-1/2})$ and
$R_0$ is densely defined.
Since $V$ is relatively form-bounded with respect to $h_0$,
$R_0^*$
is also densely defined, and
 $R_0$ is closable.
We denote the closure by the same symbol. We  define
\eq{s1}
 K_0 = R_0^* R_0.
 \en

Next let us introduce assumptions on the external potential $V$.
\begin{assumption}
\label{cc}
$V$ satisfies
that (1) $V\leq 0$ and
(2) $R_0$ is compact.
\end{assumption}

\begin{lemma}
\label{lem1339}
Suppose Assumption \ref{cc}.
Then
\begin{itemize}
\item[(i)] $R_E$, $R_E^*$ and $K_E$ ($E \leq 0$) are compact.
\item[(ii)] $\|K_E\|$ is continuous and     monotonously increasing in $E \leq 0$ and it holds that
\eq{s2}
 \lim_{E \to -\infty}\|K_E\| = 0, \quad \lim_{E \uparrow 0}\|K_E\| = \|K_0\|.
 \en
 \end{itemize}
\end{lemma}
\proof
Under (2) of Assumption \ref{cc},
$R_0^*$ and $K_0$ are compact.
Since  
\begin{equation}
\label{1240}
(f, K_E f)
 \leq (f, K_0 f), \quad f \in C_0^\infty(\RR^d),
\end{equation}
extends to $f \in L^2(\RR^3)$, $K_E$, $R_E$ and $R_E^*$ are also compact.
Thus (i) is proven.

We will prove (ii).
It is clear from \kak{1240} that $K_E$ is monotonously increasing in $E$.
Since $R_0$ is bounded, \eqref{1240} holds on $L^2(\RR^d)$ and
\begin{align}
\label{2035}
K_E = R_0^* \left((h_0-E)\f h_0\right)
R_0, \quad E \leq 0.
\end{align}
From \eqref{2035} one concludes that
$$
\| K_E - K_{E^\prime} \|
  \leq  \|K_0\| \frac{|E-E^\prime|}{|E^\prime|}
$$
for $E, E^\prime < 0$.
Hence $\|K_E\|$ is continuous in $E<0$.
We have to prove the left continuity at $E=0$.
Since $\|K_E\| \leq \|K_0\|$ ($E < 0$),
one has $ \lim\sup_{E \uparrow 0}\|K_E\| \leq \|K_0\|$.
By \eqref{2035} we see that
$K_0 = \mbox{s-}\lim_{E \uparrow 0}K_E$ and
\begin{align*}
\| K_0 f \| = \lim_{E \uparrow 0}\| K_E f \|
	\leq \left(\liminf_{E \uparrow 0}\|K_E\|\right) \|f\|, \quad f \in L^2(\RR^d).
\end{align*}
Hence we have $\|K_0\| \leq \liminf_{E \uparrow 0}\|K_E\|$
and $\lim_{E\uparrow 0}\|K_E\| = \|K_0\|$.
It remains to prove that $\lim_{E \to -\infty}\|K_E\|=0$.
Since $R_0^*$ is compact, for any $\epsilon>0$,
there exists a finite rank operator $T_\epsilon = \sum_{k=1}^n (\varphi_k, \cdot) \psi_k$
such that $n =n(\epsilon)<~\infty$, $\varphi_k, \psi_k \in L^2(\RR^d)$
and $\|R_0^* - T_\epsilon\| < \epsilon$.
Then it holds that $\| K_E \| \leq  \left(\epsilon + \|T_\epsilon h_0(h_0-E)^{-1}\|\right) \|R_0\|$.
For any $f \in L^2(\RR^d)$, we have
\[ \|T_\epsilon h_0(h_0-E)^{-1} f\|
\leq  \left( \sum_{k=1}^n \| h_0 (h_0-E)^{-1}\varphi_k \| \|\psi_k\| \right) \|f\| \]
and
 $\lim_{E \to - \infty} \|T_\epsilon h_0(h_0-E)^{-1}\| = 0$,
which completes  (ii).
\qed

Let
\eq{s10}
\hp(m) = -\frac{1}{2m}\Delta + V.
\en
By (ii) of Lemma \ref{lem1339}, we have
$\lim_{E \to -\infty}\| |V|^{1/2}(h_0-E)^{-1/2} \| = 0$.
Therefore $V$ is infinitesimally form bounded with respect to $h_0$
and $\hp(m)$ is the self-adjoint operator associated with the quadratic form
$$f, g\mapsto
\frac{1}{m}(h_0^{1/2}f,h_0^{1/2}g) + (|V|^{1/2}f, |V|^{1/2}g)$$ for
 $f, g \in D(h_0^{1/2})$.
Note that the domain $D(\hp(m))$ is independent of $m$.

Under (2) of Assumption \ref{cc}, the essential spectrum of $\hp(m)$ coincides with that of $-\frac{1}{2m} \Delta$,
hence $\s_{\rm ess}(\hp(m))=[0,\infty)$.
Next we will estimate the spectrum of $\hp(m)$ contained in $(-\infty, 0]$. Let $\one_{({\cal O})}(T)$, ${\cal O}\subset \RR$,  be the spectral resolution of self-adjoint operator $T$ and set
\eq{s13}
 N_{\cal O}(T) = \dim {\rm Ran} \one_{\cal O}(T).
 \en
 The Birman-Schwinger principle \cite{sim05}
states that
\eq{s3}
\begin{array}
{ll}
\d (E<0)
&
N_{(-\infty,\frac{E}{m}]}
\lk
\hp(m)\rk=
N_{[\frac{1}{m},\infty)}
(K_{E}),\\
& \\
\d (E=0)&
N_{(-\infty,0]}
\lk
\hp(m)
\rk
\leq
N_{[\frac{1}{m},\infty)}(K_0).
\end{array}
\en
Now let us define the constant $m_{\mathrm{c}}$ by the inverse of the operator norm of $K_0$,
\eq{mzero}
 m_{\mathrm{c}} = \|K_0\|^{-1}.
 \en
\begin{lemma}\label{schr}
Suppose Assumption \ref{cc}.
\begin{itemize}
\item[(1)] If $m < m_{\mathrm{c}}$,
then
$N_{(-\infty,0]}(\hp(m))=0$.
\item[(2)] If $m > m_{\mathrm{c}}$,
then
$N_{(-\infty,0]}(\hp(m))\geq 1$.
\end{itemize}
\end{lemma}
\proof
It is immediate to see (1)
by the Birman-Schwinger principle \kak{s3}.
Suppose
 $m > m_{\mathrm{c}}$.
 Then, using the continuity and monotonicity of $ E \to \|K\|$, see Lemma \ref{lem1339},
there exists $\epsilon > 0$ such that
$m_{\mathrm{c}}<  \| K_{-\epsilon}\|\f \leq m$.
Since $K_{-\epsilon}$ is positive and compact,
$\| K_{-\epsilon}\| \in \sigma_{\rm p}( K_{-\epsilon})$ follows and hence
$N_{[\frac{1}{m}, \infty)}
(K_{-\epsilon}) \geq 1$.
Therefore (2) follows again from
the Birman-Schwinger principle.
\qed
\begin{remark}
\label{yui}
By Lemma \ref{schr}, the critical mass at zero coupling $m_{\mathrm{c}}(0)=m_{\mathrm{c}}$.
\end{remark}
In the case $m>m_{\mathrm{c}}$,
by the proof of Lemma \ref{schr} one concludes that the bottom of the spectrum of $\hp(m)$
is strictly negative.
For $\epsilon>0$ we set
\eq{s11}
 m_\epsilon = \|K_{-\epsilon}\|^{-1}.
 \en
 \bc{schr1}
Suppose Assumption \ref{cc} and $m>m_\epsilon$.
Then
\eq{oto}
\inf \sigma\left(\hp(m) \right) \leq \frac{-\epsilon}{m}.
\en
\ec
\proof
The Birman-Schwinger principle states that
$1\leq
N_{(-\infty,-\frac{\epsilon}{m}]}\left(\hp(m)
 \right)$, since $ 1/m < \|K_{-\epsilon}\|$,
 which implies
the corollary.
\qed

\subsection{The case of the Pauli-Fierz model}
In this subsection we extend the Birman-Schwinger type estimate to the Pauli-Fierz Hamiltonian.
\bl{koko}
Suppose Assumption \ref{cc}.
If $m < m_{\mathrm{c}}$,
then the zero coupling Hamiltonian
$
\hp(m)+\hf
$
has
no ground state.
\enl
\proof
Since the Fock vacuum $\Omega$ is the ground state of $\hf$,
$\hp(m)+\hf
$ has a ground state
if and only if $\hp(m)$ has a ground state. But
$\hp(m)$ has no ground state by Lemma \ref{schr}.
Therefore
$\hp(m)+ \hf
$ has no ground state.
\qed

From now on we discuss
$U\f H(\alpha) U$ with $\alpha\not=0$.
We set
\eq{sp4}
U\f H(\alpha) U=H_0(\alpha)+W+g,
\en
where
\begin{eqnarray}
\begin{array}{ll}
\d H_0(\alpha)=\hz+ \hf,\\
\ \\
W=T\f V  T.
\end{array}
\end{eqnarray}
\bt{3}
Suppose Assumptions \ref{kimi}, \ref{1v} and \ref{cc}.
If $\mass < m_{\mathrm{c}}$,
then
$H_0(\alpha)+W+g$ has no ground state.
\et
\proof
Since $g$ is a constant,
we prove the absence of ground state of $H_0(\alpha)+W$.
Since $V$ is negative, so is $W$.
Hence
$\inf \sigma(H_0(\alpha)+W) \leq \inf \sigma(H_0(\alpha)) = 0$.
Then it suffices to show that $H_0(\alpha)+W$ has no  eigenvalues in $(-\infty,0]$.
Let $E \in (-\infty,0]$ and set
\eq{s4}
 \KK _E = |W|^{1/2}(H_0(\alpha) - E)^{-1}|W|^{1/2},
 \en
 where $|W|^{1/2}$ is defined by the functional calculus.
We shall prove now that
if $H_0(\alpha)+W$ has eigenvalue $E\in (-\infty,0]$, then $\KK _E$ has
eigenvalue $1$.
  Suppose that $(H_0(\alpha)+W-E)\varphi = 0$ and $\varphi \not=0$, then
$$\KK _E |W|^{1/2}\varphi = |W|^{1/2}\varphi.$$
Moreover if $|W|^{1/2}\varphi = 0$, then $W\varphi=0$ and hence $(H_0(\alpha)-E) \varphi = 0$, but
$H_0(\alpha)$ has no eigenvalue by Lemma \ref{koko}.
Then $|W|^{1/2}\varphi \not = 0$ is concluded
and  $\KK_E$ has eigenvalue $1$.
Then it is sufficient to see $\|\KK _E\| < 1$ to show that
$H_0(\alpha)+W$ has no eigenvalues in $(-\infty,0]$.
Notice that
$\hz $ and $T$ commute, and
\[   \left\|
\lk-\Delta
\rk^\han(H_0(\alpha)-E)^{-1}\lk-\Delta\rk^{1/2}
\right\| 
\leq 2\mass. \]
Then we have
\begin{eqnarray*}
 \|\KK _E\|
\leq
\left\|
|V|^\han \lk\hz\rk^{-\han}\right\|^2
=
\mass\|K_0\|=\frac{\mass}{m_\mathrm{c}}<1
\end{eqnarray*}
and the proof is complete.
\qed

\section{Absence and existence of a ground state}
In this section we establish the absence, resp.  existence, of a ground state of the Pauli-Fierz Hamiltonian $H_0(\alpha)+W$.
Let $\kappa>0$ be a parameter and
let us define the Pauli-Fierz Hamiltonian
with scaled external potential $V_\kappa(x) = V(x/\kappa)/\kappa^2$ by
\eq{ca}
H_\kappa
=\frac{1}{2m}(p- \alpha  A)^2+V_\kappa+\hf.
\en
We also define
$K_\kappa $ by
$H(\alpha)$ with $a^\sharp$ replaced by
$\kappa a^\sharp$.
Then
\eq{ki1}
 K_\kappa  = \frac{1}{2m}(p-\kappa \alpha A)^2 + V + \kappa^2 H_{\rm f}.
 \en
$ H_\kappa$ and $\kappa^{-2}K_\kappa$ are unitarily equivalent,
\eq{s14-1}
 H_\kappa\cong \kappa^{-2}K_\kappa.
 \en
Let $m<m_\mathrm{c}$ and $\epsilon>0$.
We define the function
\begin{eqnarray}
\label{a1}
\alpha_\epsilon  &=& (\frac{d-1}{d}\|\hat{\varphi}/\omega\|^2)^{-1/2} \sqrt{m_\epsilon -m},\quad \epsilon>0\\
\label{a2}
\alpha_0&=&  (\frac{d-1}{d}\|\hat{\varphi}/\omega\|^2)^{-1/2} \sqrt{m_\mathrm{c} - m},
\end{eqnarray}
where
we
recall that $m_\epsilon = \|K_{-\epsilon}\|^{-1}$ for $\epsilon\geq0$.
Note that
\begin{itemize}
\item[(1)] $|\alpha| < \alpha_0$ if and only if $\mass < m_\mathrm{c}$;
\item[(2)] $|\alpha| > \alpha_\epsilon $ if and only if $\mass > m_{\rm \epsilon}$.
\end{itemize}
Note that $\alpha_0 < \alpha_\epsilon $ because of $m_\epsilon > m_{\rm c}$. Since $\lim_{\epsilon \downarrow 0}m_\epsilon =m_\mathrm{c}$,
   it holds that
$  \lim_{\epsilon \downarrow 0}\alpha_\epsilon  = \alpha_0$.
We furthermore introduce assumptions
on
the external potential $V$ and ultraviolet cutoff $\vp$.
\begin{assumption}
\label{vs}
The external potential $V$ and
the ultraviolet cutoff $\vp$ satisfies:
\bi
\item[(1)]
 $V\in C^1(\BR)$ and ${\nabla V}\in L^\infty(\BR)$;
\item[(2)] $\vp/\omega^{5/2}\in\LR$.
\ei
\end{assumption}
We briefly comment on (1) of Assumption \ref{vs}.
We know that
$$H_0(\alpha)+W=-\frac{1}{2\mass}\Delta+V+\hf+
V(\cdot-\frac{\alpha}{\mass}\phi)-V.$$
The term on the right-hand side above,
$H_\mathrm{int}=V(\cdot-\frac{\alpha}{\mass}\phi)-V$,  is regarded as the interaction,
and
$$H_\mathrm{int}\sim
\frac{\alpha}{\mass}
 \nabla V(\cdot)\cdot \phi.$$
By (1) of Assumption \ref{vs},
we have
$$\|H_\mathrm{int}
 \Phi\|
\leq C \|(\hf+1)^\han\Phi\|$$
with some constant $C$ independent of $\alpha$.
This estimate follows from
the fundamental inequality $\|a^\sharp(f)\Phi\|
\leq\|f/\sqrt\omega\|
\|(\hf+1)^\han \Phi\|$.
Then the interaction has a uniform bound with respect to the coupling constant $\alpha$.
Since the decoupled Hamiltonian $-\frac{1}{2\mass}\Delta+V+\hf$ has a ground state for sufficiently large $\alpha$, it is expected that $H_0(\alpha)+W$ also has a ground state for sufficiently large $\alpha$.
This is rigorously proven in (1) of Theorem \ref{11} below.
Now we are in the position to state the main theorem.
\begin{theorem}
\label{11}
Suppose
Assumptions \ref{kimi}, \ref{1v}, \ref{cc} and  \ref{vs}.
Then (1) and (2) below hold.
\begin{itemize}
\item[(1)]
For any $\epsilon>0$,
there exists $\kappa_\epsilon$ such that
for all $\kappa >\kappa_\epsilon$,
$H_\kappa$ has a unique ground state
for all
$\alpha$ such that $|\alpha| >\alpha_\epsilon $,
\item
[(2)]
$H_\kappa$ has no ground state
for all $\kappa >0$ and
all $\alpha$ such that $|\alpha| <\alpha_0$.
\end{itemize}
\end{theorem}
\proof
Let
 $U_\kappa $ (resp. $T_\kappa $) be defined by
$U$ (resp. $T$)  with $\omega$ and $\vp$ replaced by $\kappa^2\omega$ and $\kappa\vp$.
Then
\eq{migi}
U_\kappa ^{-1}K_\kappa U_\kappa
=
\hp(\mass)
 +
   \kappa^2  H_{\rm f} +
   \delta V_\kappa
	+ g,
\en
where
$ \delta V_\kappa  = T_\kappa \f VT_\kappa - V$.
Note that $g$ is independent of $\kappa$.
Since
$U_\kappa\f K_\kappa U_\kappa $ is unitary equivalent to
$\kappa^2H_\kappa$, we prove the existence of a ground state for $U_\kappa\f K_\kappa U_\kappa$.
Let $N
= \jjj \int\add(k,j)a(k,j) dk$ be the number operator.
Since
$\hp(\mass) $ has a ground state
by the assumption $|\alpha|>\alpha_\epsilon $, i.e.,
 $\mass > m_\mathrm{c}$,
it can be shown that
$U_\kappa ^{-1}K_\kappa
U_\kappa
+\nu  N$ with $\nu>0$
also has a ground state, see \cite[p.1168]{hs01} for details.
We denote the normalized ground state of
$U_\kappa ^{-1}K_\kappa
U_\kappa
+\nu  N$ by $\Psi_\nu = \Psi_\nu(\kappa) $.
Since the unit ball  in a Hilbert space is weakly compact,
there exists a subsequence of
$\Psi_{\nu^\prime}$
such that the weak limit $\Psi = \lim_{\nu^\prime \to 0} \Psi_{\nu^\prime}$ exists.
If $\Psi\not=0$, then $\Psi$ is  a ground state \cite{AH}.
 Let   $P=\one_{[\Sigma,0)}(\hz +V)
   \otimes \one_{\{0\}}(H_{\rm f})$ and
$\Sigma=
\is(\hp(\mass))$.
Adopting the arguments in the proof of
\cite[Lemma 3.3]{hs01}, we conclude
\eq{s5}
(\Psi, P\Psi)
\geq
1- \frac{|\alpha| \varepsilon\|\vp/\omega^{5/2}\|^2}
{\kappa^3 \mass}
	- \frac{\frac{3}{2}\frac{D}{\kappa}}{\kappa^2 (|\Sigma|-\frac{3}{2}\frac{D}{\kappa})},
\en
where $\varepsilon>0$ and $D$ are constants independent of $\kappa$ and $\alpha$.
Since $\mass > m_{\epsilon} > m_{\epsilon/2}$,
\eq{yoshida}
\Sigma\leq \is(\hp(m_\epsilon))
\leq  -\frac{\epsilon}{2m_{\epsilon}}
\en
by Corollary \ref{schr1}.
By \kak{yoshida} and \kak{s5}
we have
\eq{s14}
(\Psi, P\Psi)
\geq
\kappa^{-3}\lk\rho(\kappa)-\varepsilon
 \|\vp/\omega^{5/2}\|^2\frac{|\alpha|}{\mass}\rk,
 \en
where
$\d \rho(\kappa)=\kappa^3-\frac{\kappa}
{\xi \kappa-1}$ with $\xi=\frac{2\epsilon}{3m_\epsilon D}$.
Then
there exists $\kappa_\epsilon>0$ such that the right-hand side of \kak{s14} is positive for all $\kappa>\kappa_\epsilon$ and all $\alpha\in\RR$.
Actually
a sufficient condition for the positivity of
 the right-hand side of \kak{s14} is
\eq{s15}
\rho(\kappa)>\frac
{\varepsilon\|\vp/\omega^{5/2}\|^2}{2\sqrt m\|\vp/\omega\|},
\en
since
$\sup_{\alpha} \frac{|\alpha|}{\mass}=
(2\sqrt m\|\vp/\omega\|)\f$.
Then $\Psi\not=0$
for all  $\kappa>\kappa_\epsilon$.
Thus
the ground state exists for
all $|\alpha|>\alpha_\epsilon $ and all $\kappa>\kappa_\epsilon$
and (1) is complete.

We next show (2).
Notice that
$$U_\kappa\f  H_\kappa U_\kappa=
\hz+  \hf +
T\f V_\kappa
 T+g.$$
Define the unitary operator $u_\kappa$ by $(u_\kappa f)(x) = k^{d/2} f(x/\kappa)$.
Then we infer $V_\kappa = \kappa^{-2}u_\kappa V u_\kappa^{-1}$,
$-\Delta = \kappa^{-2}u_\kappa (-\Delta) u_\kappa^{-1}$ and
\begin{align*}
\| |V_\kappa|^{1/2}(-\Delta)^{-1}|V_\kappa|^{1/2}\|
 = \kappa^{-2} \| u_\kappa |V|^{1/2} u_\kappa^{-1}(-\Delta)^{-1} u_\kappa |V|^{1/2} u_\kappa^{-1} \|
 = \| K_0\|.
\end{align*}
(2) follows from Theorem \ref{3}.
\qed

\bc{v}
Let arbitrary $\delta>0$ be given.
Then there exists
an external potential $\tilde V$ and constants
$\alpha_+>\alpha_-$
such that
\bi
\item[(1)]
$0<\alpha_+-\alpha_-<\delta$;
\item[(2)]
$H(\alpha)$
has a ground state for $|\alpha|>\alpha_+$
but  no ground state for $|\alpha|<\alpha_-$.
\ei
\ec
\proof
Suppose that $V$ satisfies Assumption \ref{cc}.
For $\delta>0$  we take $\epsilon>0$ such that $\alpha_\epsilon -\alpha_0
<\delta$.
Take a sufficiently large $\kappa$ such that \kak{s15} is fulfilled, and set $\tilde V(x)=V(x/\kappa)/\kappa^2$.
Define $H(\alpha)$ by the Pauli-Fierz Hamiltonian with potential $\tilde V$. Then $H(\alpha)$ satisfies (1) and (2) with $\alpha_+=\alpha_\epsilon $ and $\alpha_-=\alpha_0$.
\qed

\begin{remark}[Upper and lower bound of $m_\mathrm{c}(\alpha)$]
\label{kansai2}
{\rm Corollary 4.3 implies the upper and lower bounds
\eq{s16}
\begin{array}{l}
m_-(\alpha)\leq m_\mathrm{c}(\alpha)\leq
m_+(\alpha),\\
m_\mathrm{c}(0)=m_\mathrm{c},
\end{array}
\en
where
$$
\begin{array}{l}
m_-(\alpha)=m_0-\alpha^2\frac{d-1}{d}\|\vp/\omega\|^2,\\
m_+(\alpha)=m_\epsilon-\alpha^2\frac{d-1}{d}\|\vp/\omega\|^2.
\end{array}
$$
Fix the coupling constant $\alpha$.
If $m<m_-(\alpha)$, then there is no ground state, and if $m>m_+(\alpha)$, then the ground state exists,
compare with Fig. 1.
}
\end{remark}

\begin{remark}
[$m_\mathrm{c}(\alpha)$ for sufficiently large $\alpha$]
{\rm
Let
 $(\frac{d-1}{d}\|\vp/\omega\|^2)\f m_\epsilon<\alpha^2$.
 Then by Remark \ref{kansai2},
 $H(\alpha)$ has a ground state for arbitrary $m>0$.
It is an open problem to establish whether this is an artifact of the dipole approximation or in fact holds also for the
Pauli-Fierz operator.}\end{remark}


\section{Examples of external potentials}
In this section we give examples of potentials $V$ satisfying Assumption \ref{cc}.
The self-adjoint operator $h_0^{-1}$ has the integral kernel
\[ h_0^{-1}(x,y) = \frac{b_d}{|x-y|^{d-2}}, \quad d\geq 3,  \]
with $b_d = 2{\Gamma((d/2)-1)/\pi^{(d/2)-2}}$.
It holds that
\begin{equation}
\label{210}
(f, K_0f)
= \int dx \int dy \ov{f(x)}K_0(x,y)f(y),
\end{equation}
where
\eq{s12}
K_0(x,y) =
	b_d \frac{|V(x)|^{1/2}|V(y)|^{1/2}}{|x-y|^{d-2}},\quad d\geq3,
\en
is the integral kernel of operator $K_0$.
We recall the Rollnik class $\mathscr{R}$ of potentials is defined by
\[ \mathscr{R} = \left\{ V \Big| \int_{\BR} dx \int_{\BR} dy \frac{|V(x)V(y)|}{|x-y|^{2}} < \infty \right\}. \]
By the Hardy-Littlewood-Sobolev inequality,
   $\mathscr{R} \supset L^p(\RR^3) \cap L^r(\RR^3)$ with $1/p + 1/r = 4/3$. In particular, $L^{3/2}(\RR^3)\subset \mathscr{R}$.

\begin{example}{\bf  ($d=3$ and Rollnik class)}
{\rm
Let $d=3$. Suppose that $V$ is negative and $V \in \mathscr{R}$.
Then
$K_0\in
L^2(\RR^3 \times \RR^3)$.
Hence $K_0$ is Hilbert-Schmidt
and  Assumption \ref{cc} is satisfied.
}
\end{example}

The example  can be extended to dimensions $d\geq 3$.
\begin{example} {\bf ($d\geq 3$ and $V\in L^{d/2}(\BR)$)}
{\rm
Let $L_w^p(\BR)$ be the set of Lebesgue measurable function $u$ such that
$\sup_{\beta>0}
\beta\left|
\{x\in\BR||u(x)>\beta\}\right|_L^{1/p}<\infty$,
where $|E|_L$ denotes the Lebesgue measure of $E\subset\BR$.
Let $g\in L^p(\BR)$ and $u\in L_w^p(\BR)$ for $2<p<\infty$.
Define the operator $B_{u,g}$ by
$$B_{u,g}h=(2\pi)^{-d/2}\int e^{ikx} u(k)g(x)h(x)dx.$$
It is shown in \cite[Theorem, p.97]{cwi77} that
$B_{u,g}$ is a compact operator on  $\LR$.
It is known that $u(k)=2|k|^{-1}\in L_w^d(\BR)$ for $d\geq 3$.
Let $F$ denote Fourier transform on $\LR$, and suppose that $V\in L^{d/2}(\BR)$.
Then
$B_{u, |V|^{1/2}}$ is compact on $\LR$ and then
$R_0^\ast=F B_{u, V^{1/2}} F\f$ is  compact. Thus $R_0$ is also compact.
}
\end{example}

Assume that $V \in L^{d/2}(\RR^d)$.
Let us now see the critical mass of zero coupling $m_\mathrm{c}=m_0$.
By the Hardy-Littlewood-Sobolev inequality,
we have
\eq{lieb0}
|(f, K_0 f)|
\leq
D_V
\|f\|_2^2,
\en
where
\eq{tu}
D_V=
\sqrt{2} \pi
\frac{\Gamma((d/2)-1)}{\Gamma((d/2)+1)}
\lk
\frac{\Gamma(d)}{\Gamma(d/2)}\rk^{2/d}\|V\|_{d/2}^2,
\en
a constant in \kak{tu}
is proved by Lieb
\cite{lie83}.
Then
\eq{ho1}
\|K_0 \|\leq D_V.
\en
By \kak{ho1} we have
$
m_\mathrm{c} \geq D_V\f$.
In particular
in the case of $d=3$,
\eq{mz}
m_\mathrm{c} \geq \frac{3}
{
\sqrt 2\pi^{2/3}{4^{5/3}}
}
\|V\|_{3/2}^{-2}.
\en


\noindent {\bf Acknowledgments:}\\
FH acknowledges support of Grant-in-Aid for
Science Research (B) 20340032
from JSPS and
Grant-in-Aid for Challenging Exploratory Research 22654018
from JSPS.
SA acknowledges support of Grant-in-Aid for
Research Activity Start-up 22840022. We are grateful to Max Lein for helpful comments on the manuscript.
{\small

\end{document}